\documentclass{ws-procs9x6}
\usepackage{graphicx}% Include figure files
\usepackage{subfig}
\usepackage{wrapfig}
\usepackage{dcolumn}% Align table columns on decimal point
\usepackage{bm}% bold math
\usepackage{color}

\newcommand{\sqrtsNN}{\mbox{$\sqrt{s_{_{\mathrm{NN}}}}$}}
\newcommand{\dAu}{\textit{d}+Au}

\newcommand{\AuAu}{Au+Au}

\renewcommand{\AA}{\mbox{A+A}}

\newcommand{\pp}{\mbox{\textit{p}+\textit{p}}}

\newcommand{\pt}{\mbox{$p_T$}}

\newcommand{\gevc}{\mbox{${\mathrm{GeV/}}c$}}

\newcommand{\RAA}{\mbox{$R_{AA}$}}

\newcommand{\dedx}{\mbox{$dE/dx$}}

\newcommand{\Dzero}{\mbox{$D^{0}$}}

\begin{document}

\title{Heavy Flavor Production in STAR}

\author{M. Calder\'{o}n de la Barca S\'{a}nchez \\for the STAR Collaboration.\footnote{http://www.star.bnl.gov/central/collaboration/authors/authorList.php}}

\address{Physics Department, University of California\\
One Shields Ave., \\
Davis, CA 95616, USA\\
E-mail: mcalderon@ucdavis.com}

\maketitle

\abstracts{
    In this contribution, the STAR collaboration at RHIC reports
    on measurements related to heavy flavor production. We present
    results from D meson production, and from indirect reconstruction of
    heavy flavor via semi-leptonic decays, including low transverse momentum
    muons and the inclusive yield of non-photonic electrons. We focus on
    the non-photonic electrons, and present results over a broad
    range of transverse momenta ($1.2 < \pt < 10$ \gevc) in \pp, \dAu,
    and \AuAu\ collisions at \sqrtsNN\ = 200 GeV. The non-photonic
    electron yield exhibits unexpectedly large suppression in central
    \AuAu\ collisions at high \pt, suggesting substantial heavy quark
    energy loss in hot QCD matter.
}
%\pacs{13.85.Qk, 13.20.Fc, 13.20.He, 25.75.Dw}
% 13 - HEP
% 13.85.Qk          Inclusive production with identified leptons, photons, or other nonhadronic particles
% 13.20.Fc          Decays of charmed mesons
% 13.20.He          Decays of bottom mesons
% 25 - NP
% 25.75.Dw          Particle and resonance production

%Intro:
% - Light Quark RAA, azimuthal correlations
% - Heavy quarks in a hot medium (dead cone effect, heavy quark
% e-loss)
%Experiment:
% - Measuring heavy flavor in STAR
% - D reco,
% - muon reco,
% - electron reco, TOF & EMC
%Results:
% - low pt electron, D0, muon
% - electron at higher pT, spectra and R_AA
%Conclusions.

%%%%%%%%%%%%%%%%%%%%%%%%%%%%%%%%%%%%%%%%%%%%%%%%%%%%%%%%%%%%%%%%%%%%%%
% INTRODUCTION
%%%%%%%%%%%%%%%%%%%%%%%%%%%%%%%%%%%%%%%%%%%%%%%%%%%%%%%%%%%%%%%%%%%%%%
\vspace*{-1\baselineskip}
\section{Introduction}
\label{sec:Introduction}

In the study of relativistic heavy-ion collisions, one of the main
goals is to create a system of deconfined quarks and gluons in the
laboratory and to study its properties.  In the experiments
performed at the Relativistic Heavy Ion Collider (RHIC) the
expectation is that the state formed in the collision of the nuclei
will be the one described by high-temperature Quantum Chromodynamics
(QCD), the Quark-Gluon Plasma (QGP). Recent experimental studies at
RHIC have given evidence that the nuclear matter created in the
highest energy collisions exhibits properties consistent with QGP
production. In particular, it has been extablished that the matter
produced is found to be extremely opaque to the passage of hard
partons\cite{star130auau,star200auau,star200dau}, which are believed
to lose energy via gluon radiation in the dense medium before
fragmenting into hadrons\cite{gluonrad}. A better quantitative
understanding of the partonic energy loss in the medium is one of
the issues that need to be addressed.  The measurement of heavy
quark (charm and bottom) production provides key tests of the parton
energy loss mechanism \cite{Jacobs:2004qv}. Because of the large
masses of charm and bottom quarks, they are produced almost
exclusively by initial parton interactions, and their production can
be calculated by perturbative QCD \cite{pQCD}. Calculations of heavy
flavor cross-sections and spectra are available in
next-to-leading-order for both \pp\ \cite{pQCD,pQCDCalc1,pQCDCalc2}
and \AA\ collisions (including additional cold nuclear matter
effects such as initial parton scattering and shadowing
\cite{pQCDCalc2}). In this contribution, we summarize the current
status of heavy flavor measurements by the STAR Collaboration and
compare to theory.
%We discuss the direct
%D meson reconstruction in \dAu\ and \AuAu\ collisions, the
%measurement of the yield of low transverse momentum (\pt) muons
%coming from charm decay, and the non-photonic electron yield, \eeh,
%in \pp, \dAu, and \AuAu\ collisions at a nucleon-nucleon center of
%mass energy \sqrtsNN\ = 200 GeV. The data considerably extend the
%\pt\ range of earlier electron suppression studies
%\cite{Adler:2005xv}, allowing access to a phase-space where bottom
%decays are expected to be dominant. The broad kinematic coverage of
%the data significantly constrain the possible mechanisms of
%suppression in dense nuclear matter.

%%%%%%%%%%%%%%%%%%%%%%%%%%%%%%%%%%%%%%%%%%%%%%%%%%%%%%%%%%%%%%%%%%%%%%
% EXPERIMENT
%%%%%%%%%%%%%%%%%%%%%%%%%%%%%%%%%%%%%%%%%%%%%%%%%%%%%%%%%%%%%%%%%%%%%%
\vspace*{-1\baselineskip}
\section{Analysis and Results}
\label{sec:Experiment}
\begin{wrapfigure}[20]{l}[0.1\textwidth]{0pt}
    \includegraphics[width=0.38\textwidth]{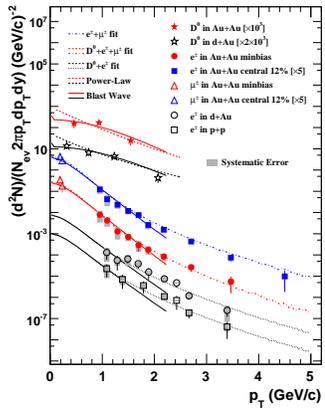}
 \caption{Open charm reconstruction summary: \pt\ distributions
for \Dzero\ mesons, for charm-decayed muons, and for non-photonic
electrons from TOF.} \label{fig:fig01}
\end{wrapfigure}
The open charm analysis relies on a combinatoric reconstruction of
the $\Dzero \rightarrow K^-\pi^+$ (and c.c.) decay chain.  The muons
originating from charm decay at low \pt\ were analyzed by combining
the energy loss (\dedx) information from the TPC and the mass-square
information from the TOF. The analysis of non-photonic electrons
consists of three main steps: selection of a clean electron sample;
subtraction of electron background arising from interactions in
material and decays; and residual corrections of the signal yield.
The electron identification was done using \dedx\ + TOF information
at low \pt, and \dedx + EMC information (matching of track momentum
and electromagnetic energy) at high \pt. The analysis details and a
discussion of the sources of uncertainty can be found elsewhere
\cite{STARDMesons1,Abelev:2006db}.

The measurements at low-\pt\ used for the estimation of the total
charm cross section are shown in Fig.~\ref{fig:fig01}. Using a
combined fit to the \Dzero, muon, and low-\pt\ non-photonic electron
spectra, we obtain the mid-rapidity \Dzero\ yield. We then
extrapolate (assuming $\sigma_{D^0}/\sigma_{c\bar{c}}=0.54 \pm
0.05$, and a factor of $4.7 \pm 0.7$ from mid-rapidity to full phase
space) to estimate $\sigma^{NN}_{c\bar{c}}$ from the
data\cite{STARDMesons1}. We obtain values in the range 0.94--1.8 mb.
The estimations agree for all datasets (as they should if
binary-collision scaling holds). The overall magnitude of the
cross-section is $\sim$ 5 times larger
 than NLO calculations, and this discrepancy is under investigation.

%Figure \ref{fig:fig01}(b) shows the inclusive (photonic and
%non-photonic) electron spectra. Corrections have been applied for
%hadron contamination, trigger efficiency, acceptance, and
%efficiency. The lines indicate the photonic spectra.
%Fig.~\ref{fig:fig01}(c) depicts the ratio of the inclusive to the
%photonic electron spectra in which a clear electron excess can be
%observed.

%%%%%%%%%%%%%%%%%%%%%%%%%%%%%%%%%%%%%%%%%%%%%%%%%%%%%%%%%%%%%%%%%%%%%%
% RESULT
%%%%%%%%%%%%%%%%%%%%%%%%%%%%%%%%%%%%%%%%%%%%%%%%%%%%%%%%%%%%%%%%%%%%%%

\begin{figure}[b]
\centering \label{fig:fig02}
    \subfloat{\includegraphics[width=0.4\textwidth]{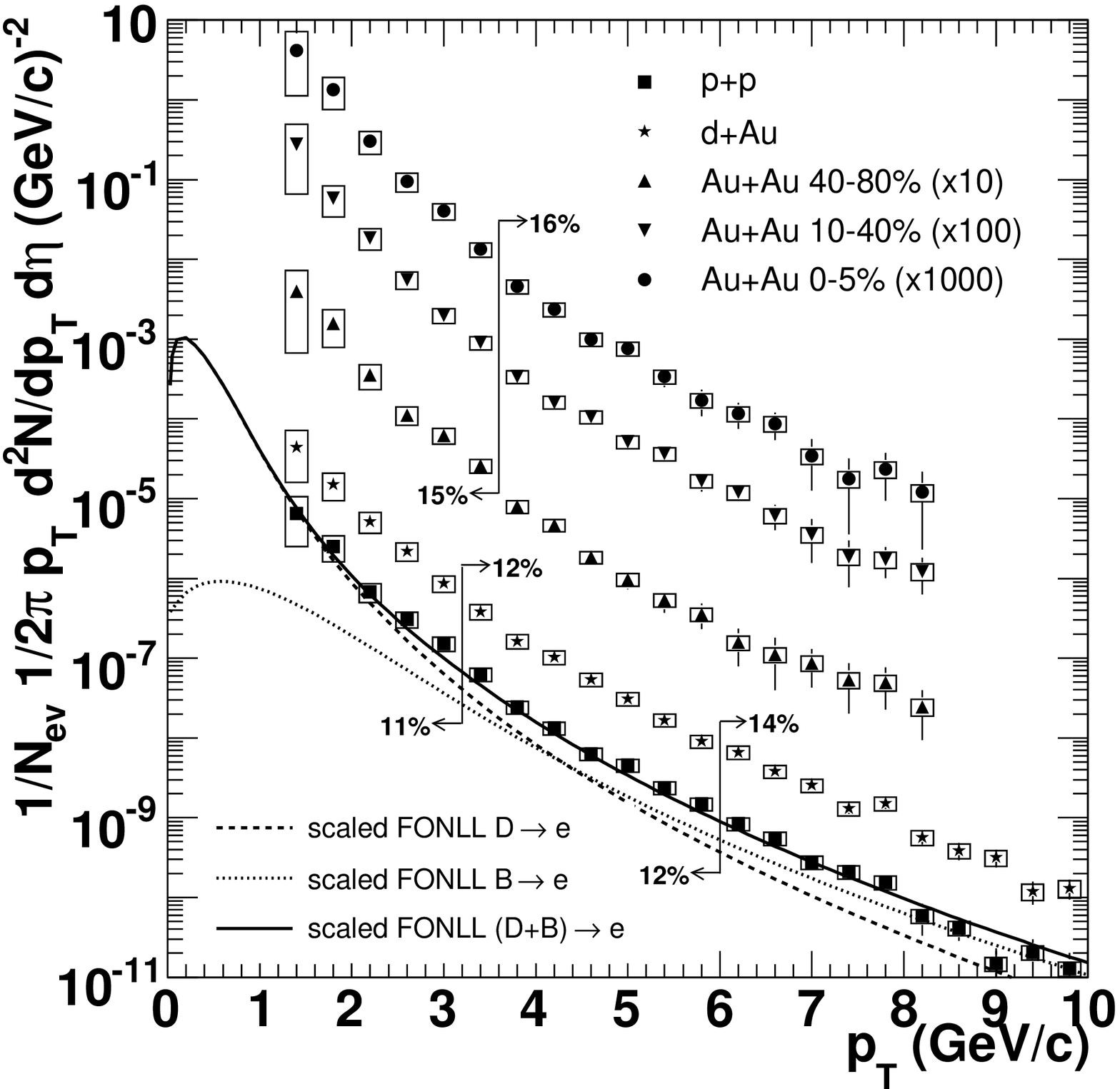}}
    \qquad
    \subfloat{\includegraphics[width=0.3\textwidth]{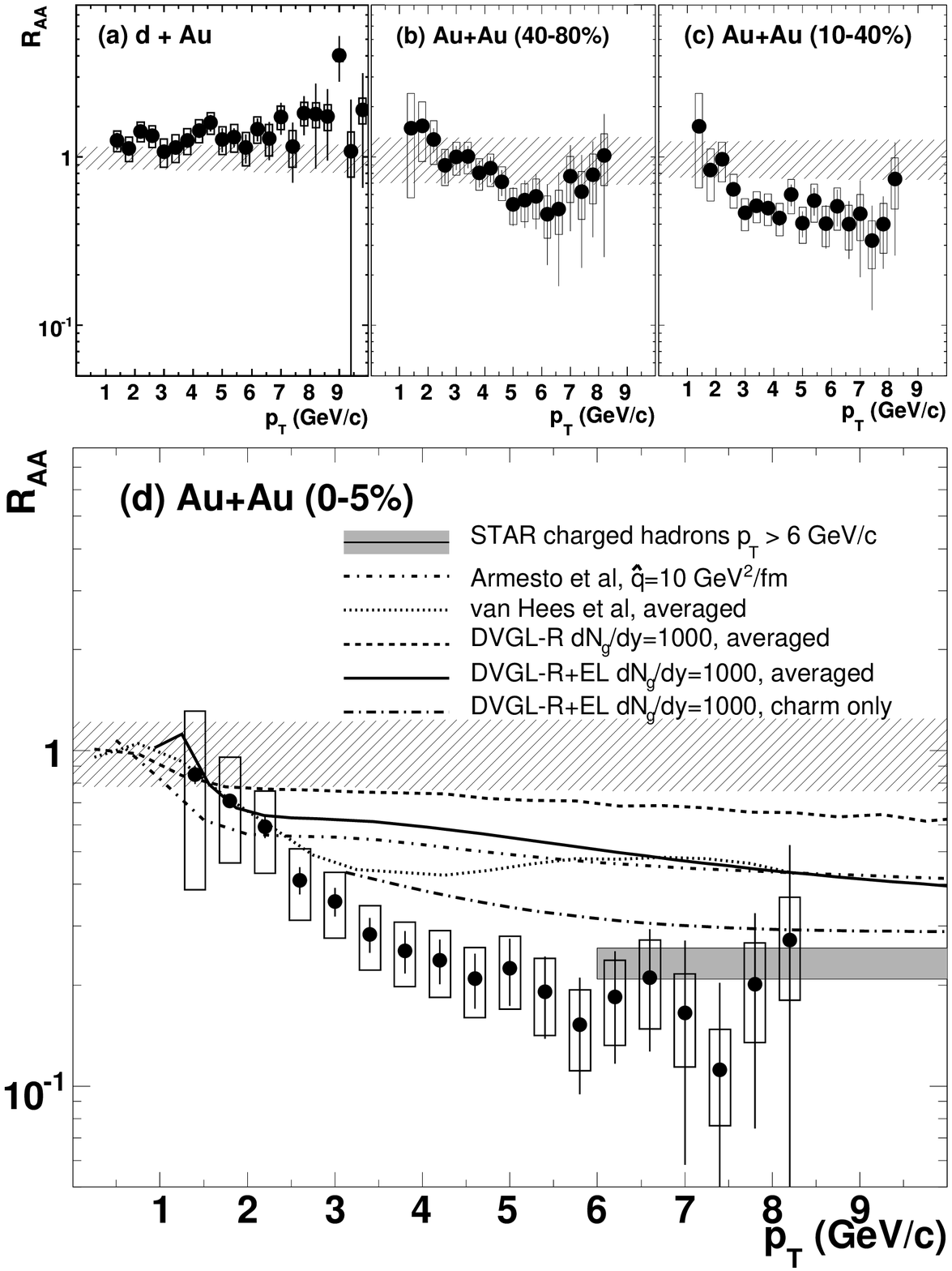}}
 \caption{Left: Non-photonic electron spectra for \pp, \dAu, and \AuAu\
  collisions at \sqrtsNN = 200 GeV.  The dashed, dotted and solid
 lines correspond to scaled pQCD predictions (see text)%\cite{pQCDCalc2}
 for the electron spectra from semileptonic
 decays of $D$ and $B$ mesons.  Right: The non-photonic electron nuclear modification factor,
\RAA, for \dAu\ and \AuAu\ collisions at
    \sqrtsNN = 200 GeV. The error bars depict the statistical
    uncertainties.}

\end{figure}

Figure \ref{fig:fig02}(left) shows the fully corrected non-photonic
electron spectra.
% for \pp, \dAu, and \AuAu\ collisions at \sqrtsNN =
%200 GeV.
%The boxes indicate the uncorrelated systematical uncertainties
%due to the corrections performed during the analysis. Table \ref{table1}
%summarizes the corrections applied to the data and their uncertainties.
%The vertical lines indicate the threshold between the minimum bias and the
%high-\pt\ electron trigger (two thresholds for \pp\ and \dAu) and the
%quoted numbers depict the overall normalization uncertainty for the
%respective dataset.
%
The dashed, dotted and solid lines correspond to pQCD predictions
\cite{pQCDCalc2} for the electron spectra from semileptonic $D$ and
$B$ mesons decays.  A common normalization factor of 5.7,
corresponding to the ratio between STAR measured charm cross section
\cite{STARDMesons1} and the FONLL cross section \cite{pQCDCalc2} is
applied to the predicted electron spectra. The calculations describe
the shape of the measured spectra rather well, although the
uncertainties in the theory and the data do not allow for a precise
determination of the region where bottom decays start to dominate.
It appears, however, that the non-photonic electrons have a
significant to dominant contribution from beauty decays at higher
\pt. Figure \ref{fig:fig02}(right) shows \RAA\ for non-photonic
electrons as a function of \pt. The error bars correspond to the
statistical uncertainties. The boxes represent the uncorrelated
systematic uncertainties while the dashed area shows the overall
normalization uncertainty. In \AuAu\ collisions, we observe an
unexpectedly strong suppression increasing from peripheral to
central collisions. For the 0-5\% most central collisions,
non-photonic electron production for $\pt > 3$ \gevc\ is suppressed
by a factor $\sim 5$ (similar to the one observed for inclusive
hadrons \cite{star200auau}, shown in grey box).

Figure \ref{fig:fig02}(right) also shows different theoretical
predictions for suppression in central events. While all depicted
calculations \cite{Djordjevic:2005,Armesto,Wicks:2005,Hess:2005} are
based on the ansatz that heavy quarks lose their energy due to final
state interactions, the predictions differ in the processes and
mechanisms taken into account.
%For example, we show calculations
%applying radiative energy loss (dashed \cite{Djordjevic:2005},
%dash-dotted\cite{Armesto}); radiative \textit{and} elastic energy
%loss, (solid curve \cite{Wicks:2005}). Other authors
%\cite{Hess:2005} (dashed-dotted curve) focus on elastic scattering
%of heavy quarks in the medium mediated by resonance excitations ($D$
%and $B$) off light quarks.
The main message is that all current models overpredict \RAA\ at
high-\pt. It is important to note that in all calculations charm
quarks are substantially more quenched than bottom quarks. The
calculated \RAA\ for electrons solely from $D$ decay describes the
data rather well. It is the dominance of electrons from $B$ decays
for $\pt \gtrsim 4$ \gevc\ that pushes the predicted \RAA\ to higher
values. The question of whether or not this discrepancy indicates
that the $B$ dominance sets in at higher \pt\ remains open until we
are experimentally able to disentangle $B$ and $D$ contributions.

%%%%%%%%%%%%%%%%%%%%%%%%%%%%%%%%%%%%%%%%%%%%%%%%%%%%%%%%%%%%%%%%%%%%%%
% SUMMARY
%%%%%%%%%%%%%%%%%%%%%%%%%%%%%%%%%%%%%%%%%%%%%%%%%%%%%%%%%%%%%%%%%%%%%%
\vspace*{-1\baselineskip}
\section{Summary and Conclusions}

We summarized the STAR results on heavy flavor production in
heavy-ion collisions. We presented measurements on direct open heavy
flavor production and on semileptonic (muon and electron) decays of
open heavy flavor leading to an estimate of the charm cross section
at RHIC energies. The nuclear modification factor of non-photonic
electrons at high-\pt\ indicates an unexpectedly large suppression
in central \AuAu\ collisions, consistent with substantial energy
loss of heavy quarks in the medium. Although all the model
calculations overpredict the data, it is important to keep in mind
that there are significant uncertainties in the data as well as in
the current calculations.  These measurements provide constraints
for models driving to a full understanding of the energy loss
mechanisms, a fundamental milestone for the characterization of the
medium properties.

\end{document}